\documentclass[a4paper,11pt]{article}
\pdfoutput=1 

\usepackage{jinstpub} 
\usepackage{lineno}

\title{\boldmath A method of detector and event visualization with Unity in JUNO}

\author[a]{Jiang Zhu,}
\author[a]{Zhengyun You,}
\author[a,1]{Yumei Zhang,\note{Corresponding author.}}
\author[a]{Ziyuan Li,}
\author[a]{Shu Zhang,}
\author[b]{Tao Lin,}
\author[b]{Weidong Li}

\affiliation[a]{Sun Yat-Sen University, Guangzhou 510275, China}
\affiliation[b]{Institute of High Energy Physics, Chinese Academy of Sciences, Beijing 100049, China}

\emailAdd{zhangym26@mail.sysu.edu.cn}

\abstract{
A visualization method based on Unity engine is proposed for the Jiangmen Underground Neutrino Observatory (JUNO) experiment.
The method has been applied in development of a new event display tool named ELAINA (Event Live Animation with unIty for Neutrino Analysis), which provides an intuitive way for users to observe the detector geometry, to tune the reconstruction algorithm and to analyze the physics events.
In comparison with the traditional ROOT-based event display, ELAINA provides better visual effects with the Unity engine.
It is developed independently of the JUNO offline software but shares the same detector description and event data model in JUNO offline with interfaces.
Users can easily download and run the event display on their local computers with different operation systems.}

\keywords{Neutrino detectors, Software architectures (event data models, frameworks and databases), Image filtering}

\begin{document}
\maketitle
\flushbottom

\section{Introduction}\label{sec:1}

In modern High Energy Physics (HEP) experiments, the detector has become more and more complex.
It is difficult for collaborators to check the detector structure and elements, as well as to understand the physics events that happen in it.
Therefore, the method of detector and event visualization is important for HEP experiments.
Good visualization helps the physicists to learn about the detector geometry, the performance of simulation and reconstruction, and the tracks and hits distributions when they are doing physics analysis or tuning the algorithm in software development. 
The visualization application of an experiment is usually called event display.

Unity is a well-known game engine \cite{bib:Unity}. It is famous for its support of multiple platforms, including the desktop, mobile device, web and Virtual Reality (VR).
Unity has been widely used in commercial game development because the developers can easily build their projects into different platforms with Unity.
The personal edition of Unity is free for people who do not make sizable profits with it and the personal edition includes most of the functions of Unity.
Visualization in HEP is basically to display the appearance of detector elements and to realize their changes with time and user interactions.
The game engine is perfectly suitable to build a visualization tool in HEP experiments as a game is essentially to display the 3D models and to respond to the actions of players. 
Based on the features of the Unity engine, we can apply it for visualization in HEP experiments and expect better performance.

\section{Event display and Unity}\label{sec:2}

ROOT \cite{bib:ROOT} is a popular data analysis framework in HEP.
It has been widely used in many HEP experiments. 
As physicists used to process data and do physics analysis with ROOT, the event display software in HEP experiments are usually built with ROOT.
The Event Visualization Environment (EVE) \cite{bib:EVE} package of ROOT provides a convenient framework for development of event display programs in HEP experiments.
It was firstly developed in the ALICE offline project \cite{bib:ALICE} and has later been used in other experiments. 
For example, one of the event display tools for the CMS \cite{bib:CMS} experiment, Fireworks \cite{bib:Fireworks}, is developed with the EVE package.
An experiment may have multiple event display tools to meet different user requirements. 
For instance, Atlantis \cite{bib:Atlantis} and VP1 \cite{bib:VP1} are the two general-purpose event display tools in the ATLAS \cite{bib:ATLAS} experiment. 
Not all the event display tools are built from scratch with 3D graphic libraries. 
Some of them are developed on top of the existing applications, such as the visualization system with SketchUp \cite{bib:SketchUp} in CMS.

Jiangmen Underground Neutrino Observatory (JUNO) is a next-generation reactor neutrino experiment. 
JUNO is currently under construction in Jiangmen city, Guangdong province, China.
It is located in a distance of 53km from both of Yangjiang and Taishan Nuclear Power Plants (NNP) \cite{bib:JUNO_CDR} and will capture the neutrinos generated by the two NPPs.
The main scientific purpose of JUNO is to determine the mass hierarchy of the three generation neutrinos by measuring the fine oscillation in observed neutrino spectrum \cite{bib:JUNO_physics}.

Software for Event display with ROOT Eve in Neutrino Analysis (SERENA) \cite{bib:SERENA} is the original event display software in the JUNO experiment, which is built with ROOT EVE and is intergraded into the JUNO offline software.
The offline software system is usually set up in the server with the operation system of Scientific Linux. 
For communications between the server and user clients, additional software such as XQuartz \cite{bib:XQuartz}, Virtual Network Console (VNC) \cite{bib:VNC} or TeamViewer \cite{bib:TeamViewer} are needed. 
However, the software cannot perfectly solve the remote display problem for all situations. 
Some of them need the administrator privileges to set up and the others can only receive the graphic window of some programs.
The network connection can also be a problem. 
Different from the situations when users receive the histograms remotely while analyzing data, the event display program has much higher requirements for network connection due to the real-time interactions. 
If the network latency between server and the client is too high, the quality of user experience in visualization and interactions will be greatly reduced. 
To avoid the problems caused by remote network connection, another idea is to setup the event display software locally. 
However, due to the strong dependence on ROOT and offline software framework, users have to install the whole offline software to run the event display. 
Many offline software of HEP experiments is only available in specific platforms such as Linux, but not so friendly in MacOS, Windows or other mobile platforms.

Considering the strong ability of multi-platform of the Unity engine, it is worth applying the Unity engine into the visualization of HEP experiments. 
Unity is a game engine and what a game engine need to do is to display elements and build reactions. 
For example, in a race game, the cars are built into a scene and can be controlled by the players when they are pressing some buttons. 
This is the same for detector and event visualization in HEP experiments. 
First, every detector elements need to be built and displayed. 
Then the users should be allowed to interact with the detector and events, such as to select specific detector elements, to rotate the view and to play the event animations with the input devices such as mouse.
Therefore, a game engine is suitable to build a visualization system.
Unity is not only used for game development, but also has many successful applications in education, simulation and visualization. 
One example in HEP is the Cross-platform Atlas Multimedia Educational Lab for Interactive Analysis (CAMELIA) \cite{bib:CAMELIA}, which is built for exploration of the ATLAS detector and the events in it. 

The motivation to propose a method based on Unity for detector and event visualization is to improve the performance and user experience in HEP experiments. 
Unity has three obvious advantages in development of the event display software, as described below.

First, as a professional game engine that has been widely used in industry, Unity can provide much fancier visual effects than most of the software in HEP such as ROOT can do. 
The periodic updates of the Unity engine also help the visualization in HEP to keep up with the latest visualization technology from industry.

Second, the method makes it easy to download and run the event display software in local machines without installation of the whole offline software. 
It can work well in some bad internet connection situations.
Instead of running the event display program on a server which usually has no GPU cards and can only provide limited graphic display power, the users can make full use of the GPU card installed on their local computers, if there is any, to accelerate the graphic display and to improve the visualization performance, especially when there are hundreds of  thousands of units to display at the same time. 

Third, Unity has good supports for multiple platforms. 
Besides the most commonly used Linux system in HEP, the applications built with Unity can be easily deployed on MacOS, Windows, web browsers, mobile devices such as pads and cellphones, and even Virtual Reality devices, which makes the event display more widely used and more convenient for professional work of physicists and public outreach. 

\section{Software structure and data flows}\label{sec:3}

\begin{figure}[htbp]
\centering 
\includegraphics[width=14cm]{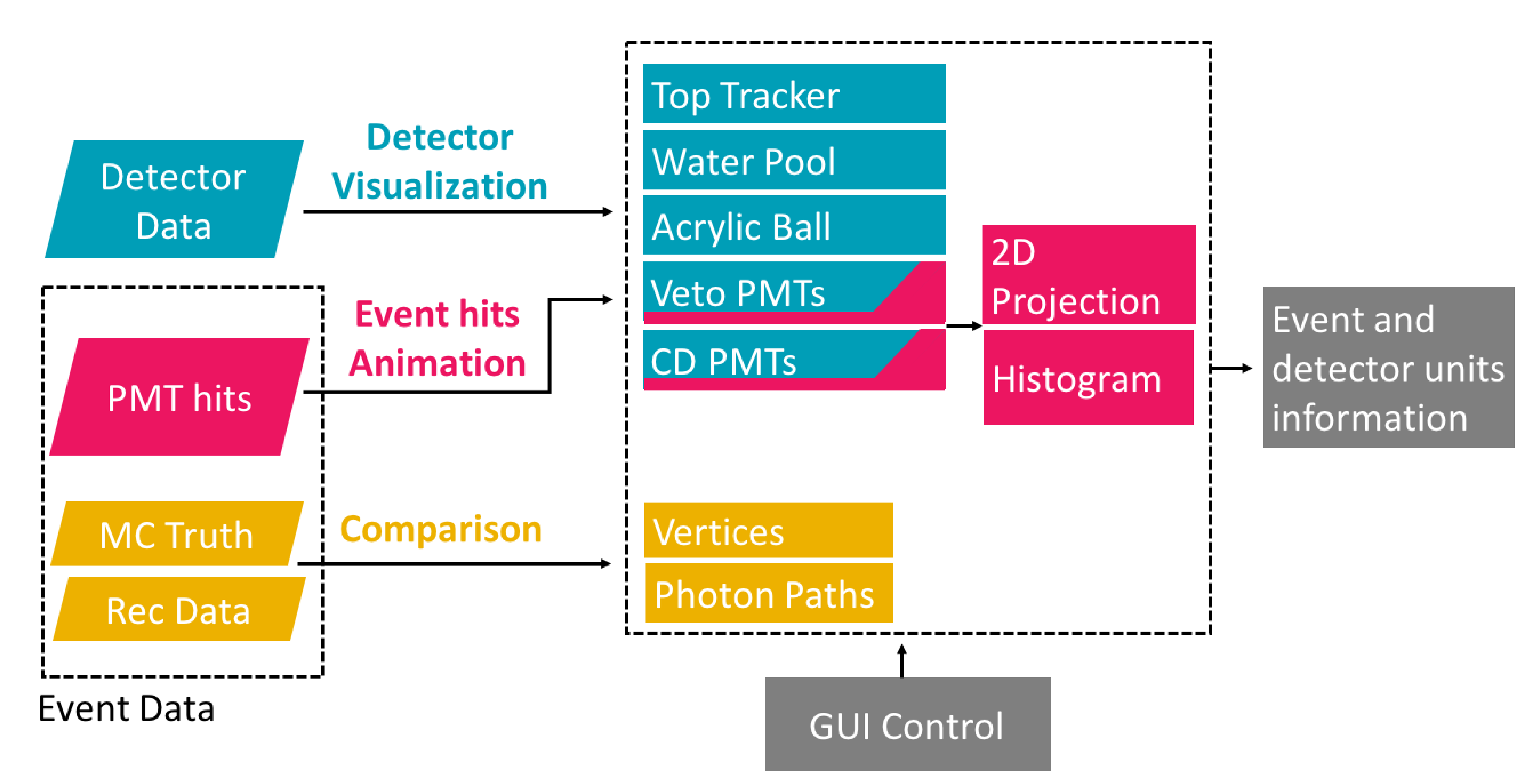}
\caption{\label{fig_dataFlow} Structure and data flows of the event display software ELAINA.}
\end{figure}

A visualization system for HEP experiments should be able to provide the functions to visualize the detector structure, to animate the event tracks and hits distribution, and to show important information, such as the details of detector hits and event information.
The Unity based visualization has been applied in the JUNO experiment to develop a new event display software, which is named ELAINA (Event Live Animation with unIty for Neutrino Analysis).
The structure and data flows of ELAINA is shown in Fig.~\ref{fig_dataFlow}.

The structure of ELAINA is composed of three parts. 
First, the detector data will be loaded and be used to construct the 3D detector objects in the scene of Unity, such as the central acrylic sphere, the Photomultiplier Tubes (PMT) \cite{bib:PMT}, the Water Pool (WP) and the Top Tracker (TT). 
Details of the JUNO detector will be introduced in the next section.
Second, the event data generated by JUNO offline software will be read in and be visualized with the objects constructed in the first step. 
The detailed information of a specific PMT, such as the number of hits on it, time of hits and total charges, will be attributed to the corresponding objects so that the users can check the information when necessary. 
The event data will be fed into histograms and be projected into 2D plots for good illustration of the event. 
Third, the event reconstruction results and the Monte Carlo truth information (for simulation data only) will be loaded to generate the reconstruction vertex and the true position of energy deposit  for comparison. 
The true path of optical photons in propagation can also be imported from simulation data to be visualized for detailed analysis of the simulation event. 
Finally, with the Graphic User Interface (GUI) control, users can modify the display effects to get a better visualization of the detector and event.

The geometry and event data are two necessary inputs to run the visualization system. 
The geometry data describing the detector structure will be used to construct the detector. 
For JUNO experiment, the original geometry information is stored as text files. 
When running the detector simulation based on Geant4 \cite{bib:Geant4}, the geometry text files will be loaded to construct detector and be converted into Geometry Description Markup Language (GDML) file \cite{bib:GDML} for persistency. 
GDML is an application-independent geometry description data format. 
The detector data exported by simulation is used as the default detector data input in ELAINA. 
The event data will be generated by users in JUNO offline software. 
Usually, the output event data files of simulation, calibration and reconstruction are all in ROOT format. 
However, the event data files cannot be read directly outside the offline software because of their specific event data models \cite{bib:EDM}. 
A macro in JUNO offline is provided to extract the event data into the readable text format for ELAINA. 

\section{Visualization and performance}\label{sec:4}
\subsection{Detector}
Visualization of the detector and its structure in 3D space is an important part of event display.  
The detectors in modern HEP experiments usually have complex structure with hundreds of thousands of units. 
A good visualization of the detector can help physicists to understand its hierarchy and spatial distribution of the detector elements precisely. 

In ELAINA, there are two ways to initialize the detector geometry in Unity.
One is to import the geometry model files directly. 
Unity supports two types of model file formats, exported 3D files and proprietary 3D application files. 
It means that Unity can directly load the geometry files that are generated by some 3D modeling software such as Blender \cite{bib:Blender}, Maya \cite{bib:Maya} and 3ds Max \cite{bib:3ds}, or load the generic formats exported by other applications.
Some HEP experiments use GDML to persistently store the detector geometry data \cite{bib:GDMLMethod}\cite{bib:BesGDML}. 
There are several open-source projects like CADMesh \cite{bib:CADMesh} and FreeCAD with GDML module \cite{bib:FreeCAD} providing the conversion between GDML and CAD files to initialize the detector.
This is the ideal way for detector initialization. 
The full detector structure will be loaded and the visualization will precisely match the detector description in offline software. 
However, associating the detector elements to the event data will be difficult in this way, because the association between the detector element and its corresponding detector identifier is lost during geometry data transformation. 
This method is also heavily dependent on the availability of transformation software.
 
The other way to initialize detector geometry is to extract the necessary information from detector data to construct the detector with scripts in Unity. 
The GDML files can be converted into ROOT format, which is easy for human to learn about the detector hierarchy with ROOT TBrowser \cite{bib:ROOT}.  
The basic information of detector elements like positions, directions, shapes, sizes and identifiers can be extracted. 
The Unity-based visualization system will create the corresponding detector objects in 3D space with such information.
This is a simple way for initialization because it only uses the necessary data to realize the detector visualization and does not depend on any external transformation tool.
It is direct and reliable when the hierarchy of the detector structure is not too complex. 
With the mapping between every detector object and its identifier being saved, the associations between detector elements and event data are well kept.  
However, different from the way of importing detector geometry automatically, it requires the developers to have good knowledge of the detector geometry before data extraction. 
If the hierarchy of the detector is very complicated, the extraction process will be difficult.

\begin{figure}[htbp]
\centering 
\includegraphics[width=12cm]{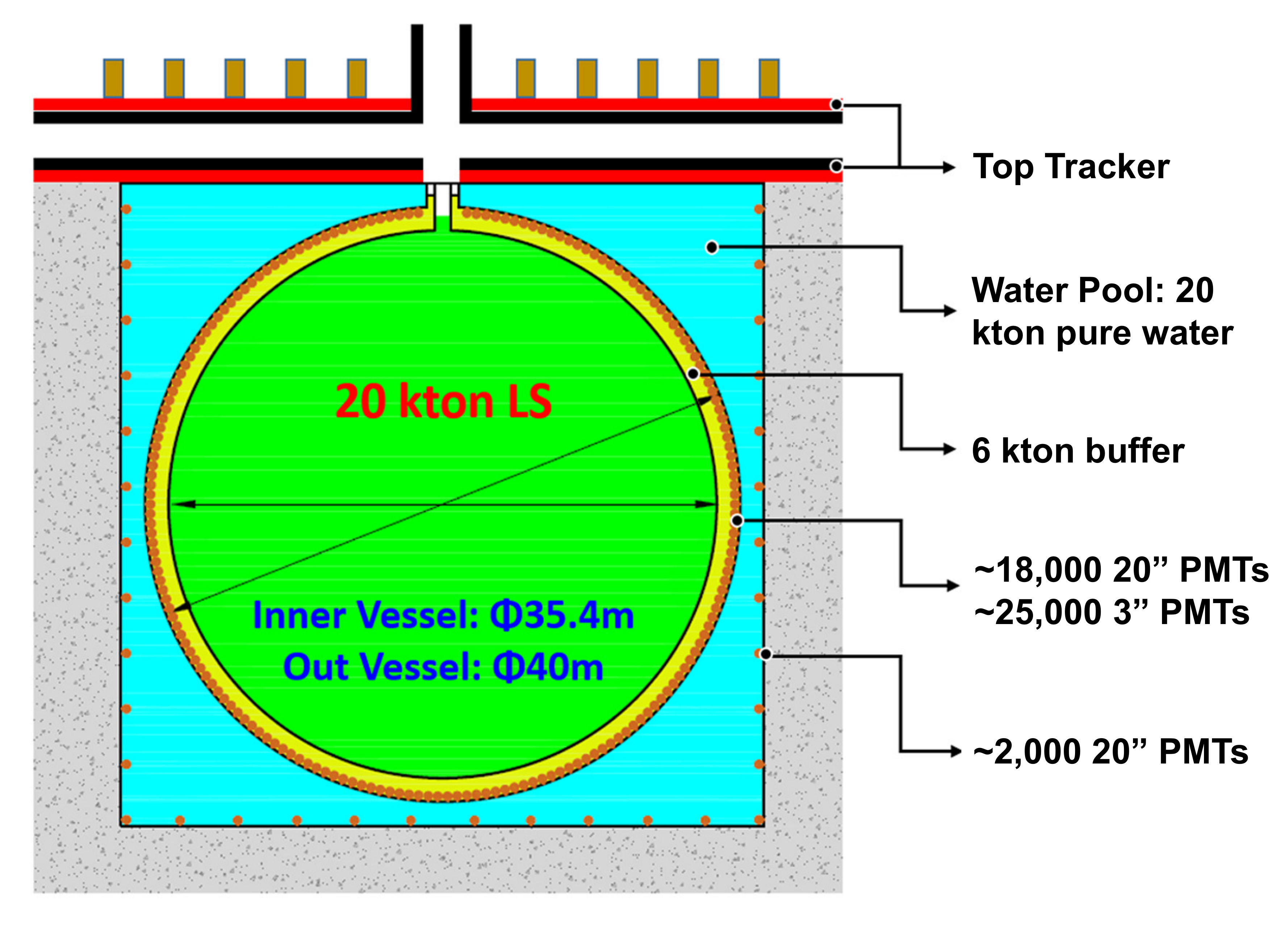}
\caption{\label{fig_structure} Structure of the JUNO detector \cite{bib:JUNO_CDR}. 
}
\end{figure}

With the detector data exported by the geometry service \cite{bib:JUNO_Geometry} in JUNO offline software, visualization of the detector can be quickly implemented with some basic 3D models using the shapes of cube, sphere and cylinder.
The structure of JUNO detector \cite{bib:JUNO_CDR} is shown in Fig.~\ref{fig_structure}.
There are about 18,000 20-inch PMTs on the surface of Central Detector (CD), which is a 35.4-meter diameter acrylic sphere with 20k tons of  Liquid Scintillator (LS) inside.
The Water Pool (WP) with 2,000 PMTs and the Top Tracker (TT) above the CD are used to veto the background particles. 

\begin{figure}[htbp]
\centering 
\includegraphics[width=12cm]{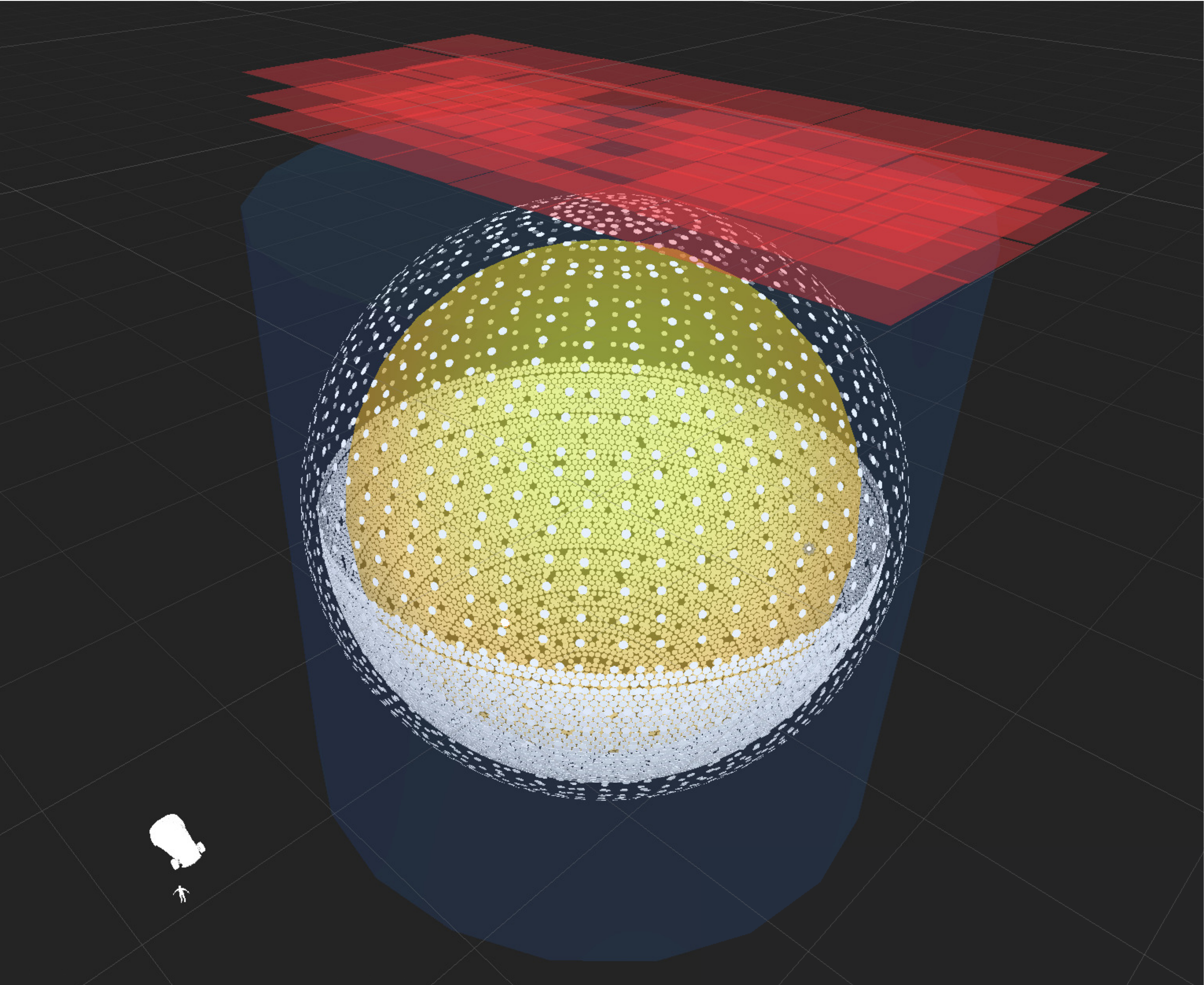}
\caption{\label{fig_detector} Visualization of the JUNO detector. 
The top red layers are plastic scintillator bars in Top Tracker. 
The small white points are 20-inch PMTs in Central Detector and Water Pool. 
The two models in the bottom left corner show the true size of a car and a human.}
\end{figure}

Fig.~\ref{fig_detector} shows the simple mode of the JUNO detector after its implementation in the Unity-based visualization system ELAINA. 
On the top of the detector, the red layers are the Top Tracker with plastic scintillators. 
The yellow sphere is the acrylic sphere in the Central Detector. 
The arrangement of CD PMTs and WP veto PMTs, which are shown as small white points, can be seen clearly. 
For better view of the CD structure, only the bottom half of 20-inch PMTs in CD are shown. 
Since the total number of PMTs in JUNO is more than 20,000, to guarantee smooth display of the whole JUNO detector, the shape of each PMT is simplified into a small cylinder, which looks like a small white point in the figure.

\begin{figure}[htbp]
\centering 
\includegraphics[width=12cm]{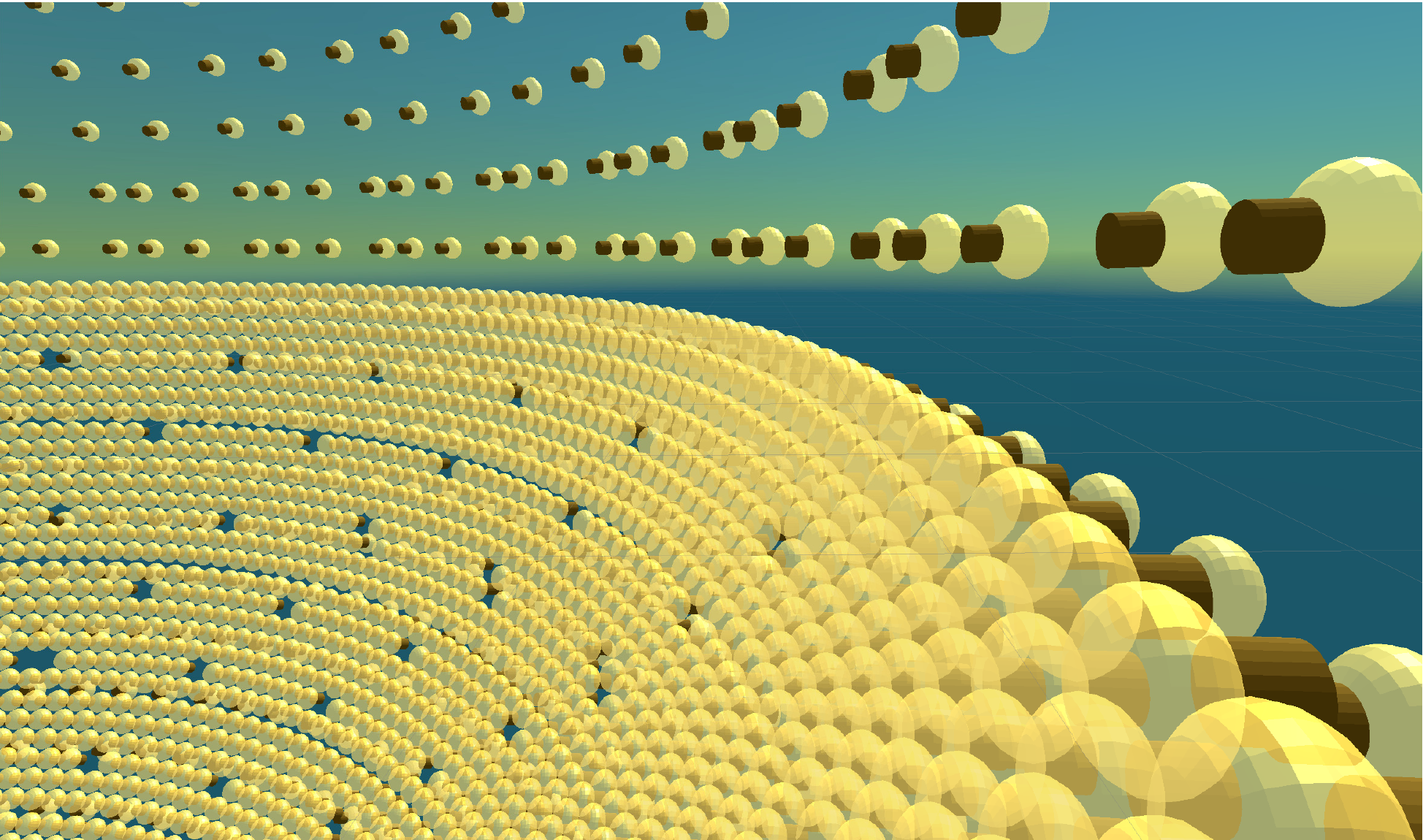}
\caption{\label{fig_pmt} Visualization of an area in the JUNO detector with the true PMT shape switch on. 
}
\end{figure}

If the users want to view the true shape of the PMTs, another full visualization mode is provided and can be easily switched to.
As shown in Fig.~\ref{fig_pmt}, the true shape of every PMT will be drawn instead of the simple cylinder. 
Hence the total number of vertices drawn in the scene will increase. 
Correspondingly, the frame refreshing rate will drop while running the event display. 
Each type of element at sub-detector level has its own switch between the simple mode and the true shape mode, so that the users have the freedom to determine whether to show the details of some detector elements if they are interested or hide for better performance.

\subsection{Event hits}
After construction of the detector in 3D space, the event hits can be displayed. 
This process is to reproduce the moment when the detector is triggered by particles in an event. 
There are a number of sensitive detectors in HEP experiments, such as the PMTs in JUNO. 
When the detector element receives a signal, its hit status will be recorded, such as the status of a PMT whether it is fired or not. 
From the spatial distribution of all sensitive detectors and their hit status in an event, the vertices or tracks of the particles can be reconstructed by reconstruction algorithm. 
Then the users can get physics information of a track such as its energy.

\begin{figure}[htbp]
\centering 
\includegraphics[width=12cm]{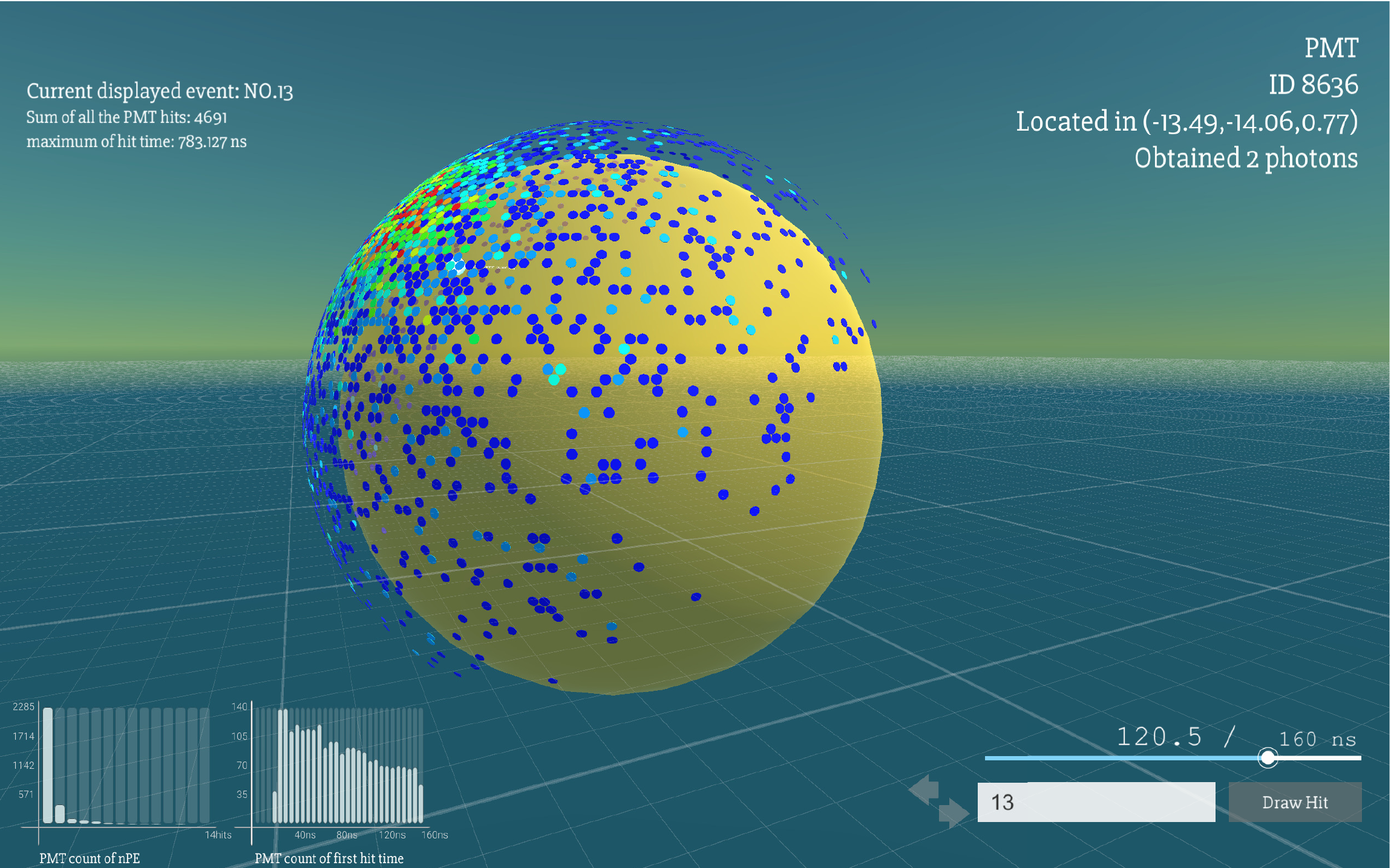}
\caption{\label{fig_hits} The main panel of ELAINA. Hits distribution of an event is shown in the center. The event control and detailed information of the event are shown in the four corners.}
\end{figure}

The typical energy of reactor neutrinos detected by JUNO is at several MeV level.
Neutrinos hit the Liquid Scintillator (LS) with IBD interactions and produce positrons and neutrons, which deposit their energy in the LS and yield photons. 
The photons propagate in the LS and finally reach the PMTs to produce hits.
By running a macro in JUNO offline software, users can extract the PMT hits data from the generated event data files. 
The hits data is converted into text format for the Unity-based visualization system to read directly. 
The identifier of the detector elements will be retained to map the PMT hits with their associated 3D detector objects in the Unity scene.
Fig.~\ref{fig_hits} shows the hits distribution of a positron event in JUNO Central Detector. 
The colorful points on the acrylic sphere are fired PMTs in the event and the different colors represent different number of hits on each PMT. 
The animation can be played to show how the detector reacts when receiving the signals. 
The users can also adjust the time range to observe the hits distribution at specific moments, which gives a intuitive illustration of the event evolution with time.

\subsection{Simulation and reconstruction comparison}
At the stage of detector design and construction, Monte Carlo (MC) simulation is a very useful tool to optimize the detector structure.
It helps physicists to test the performance of the detector and to improve the algorithm in offline software. 
The MC simulation also plays an important role in predicting whether a specific detector design can meet the requirements of the physics goals of an experiment.

One of the most important functions of event display is to help tuning reconstruction algorithms.
Reconstruction means using the signals received by the detector to restore the true physics information as closely as possible. 
For simulation case, it is to reconstruct what Monte Carlo simulation has generated, i.e. the MC truth.
By comparing the reconstruction results with corresponding MC truth information in event display, the software developers can check the accuracy of reconstruction algorithm, diagnose the reconstruction results event by event to find the potential problems and to improve the reconstruction algorithm.

\begin{figure}[htbp]
\centering 
\includegraphics[width=12cm]{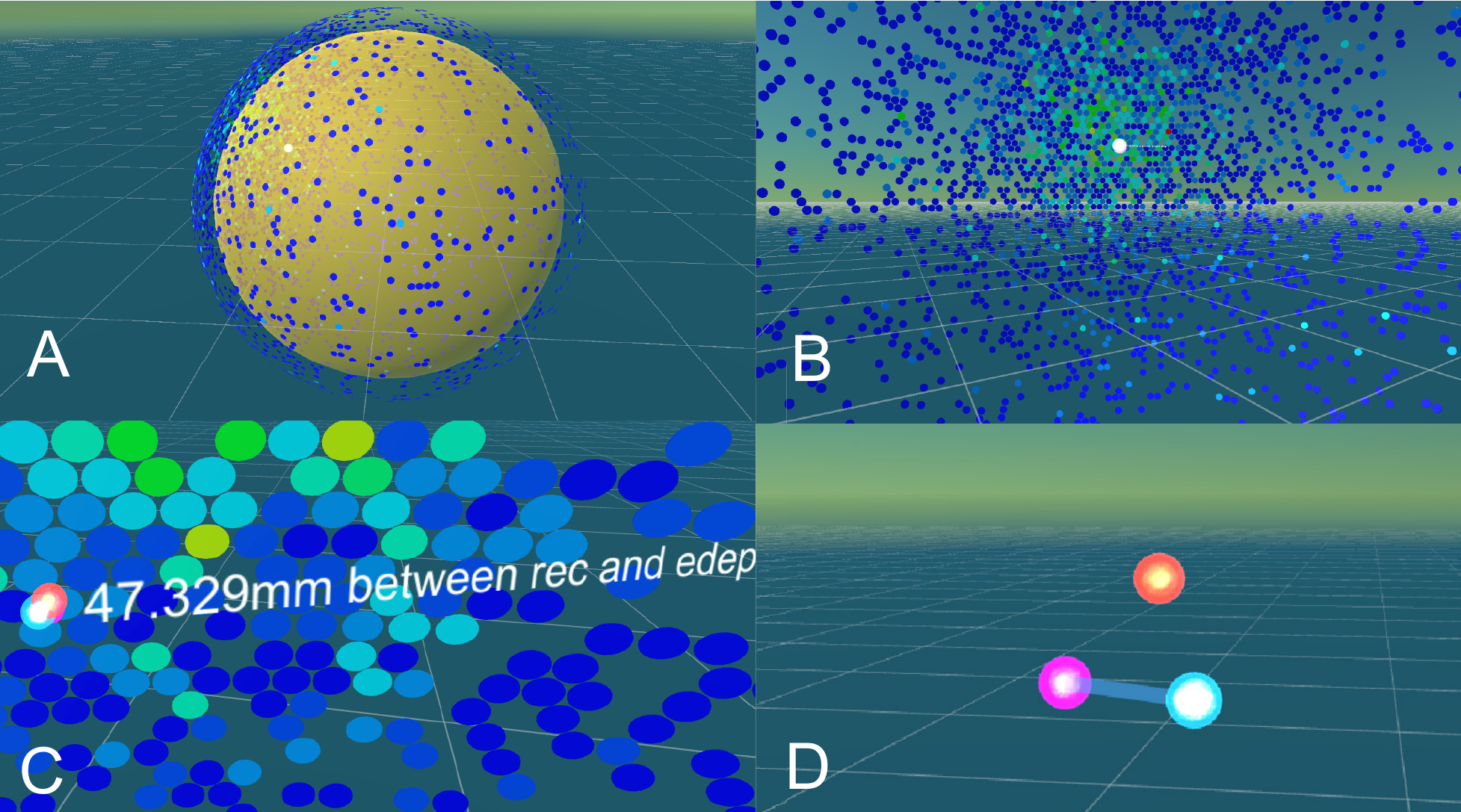}
\caption{\label{fig_comparison} Comparison of the reconstructed vertex with MC truth in a positron event. A, B, C and D give the zoom-in view of the vertex area step by step. 
The red, blue and purple points represent the particle's initial production vertex, the true energy deposit vertex and the reconstructed vertex, respectively.
The initial production vertex is where the positron is generated. 
The true energy deposit vertex is the point where the positron annihilates and deposits its energy to excite photons in the LS.}
\end{figure}

JUNO aims at precision measurement of the reactor neutrino energy spectrum for mass hierarchy determination, so it is critical to reconstruct the interaction vertex and energy of each neutrino event as precisely as possible.  
The MC truth and reconstruction results can be exported together with the PMT hits when extracting the event data files, such as the true energy of a particle, its energy deposit vertex, the reconstructed event vertex and energy, as well as the true optical photon paths in propagation. 
Fig.~\ref{fig_comparison} shows the comparison between the reconstructed vertex and MC truth in ELAINA. 
A straight line connecting the two points makes them easier to be found.
It also tells the users the performance of reconstruction.
A shorter distance between the true and reconstructed vertices usually indicates that the event vertex is better reconstructed.
Since the neutrino signal events are usually very rare, ELAINA will be especially useful in event scanning and hand selection in JUNO.

\begin{figure}[htbp]
\centering 
\includegraphics[width=12cm]{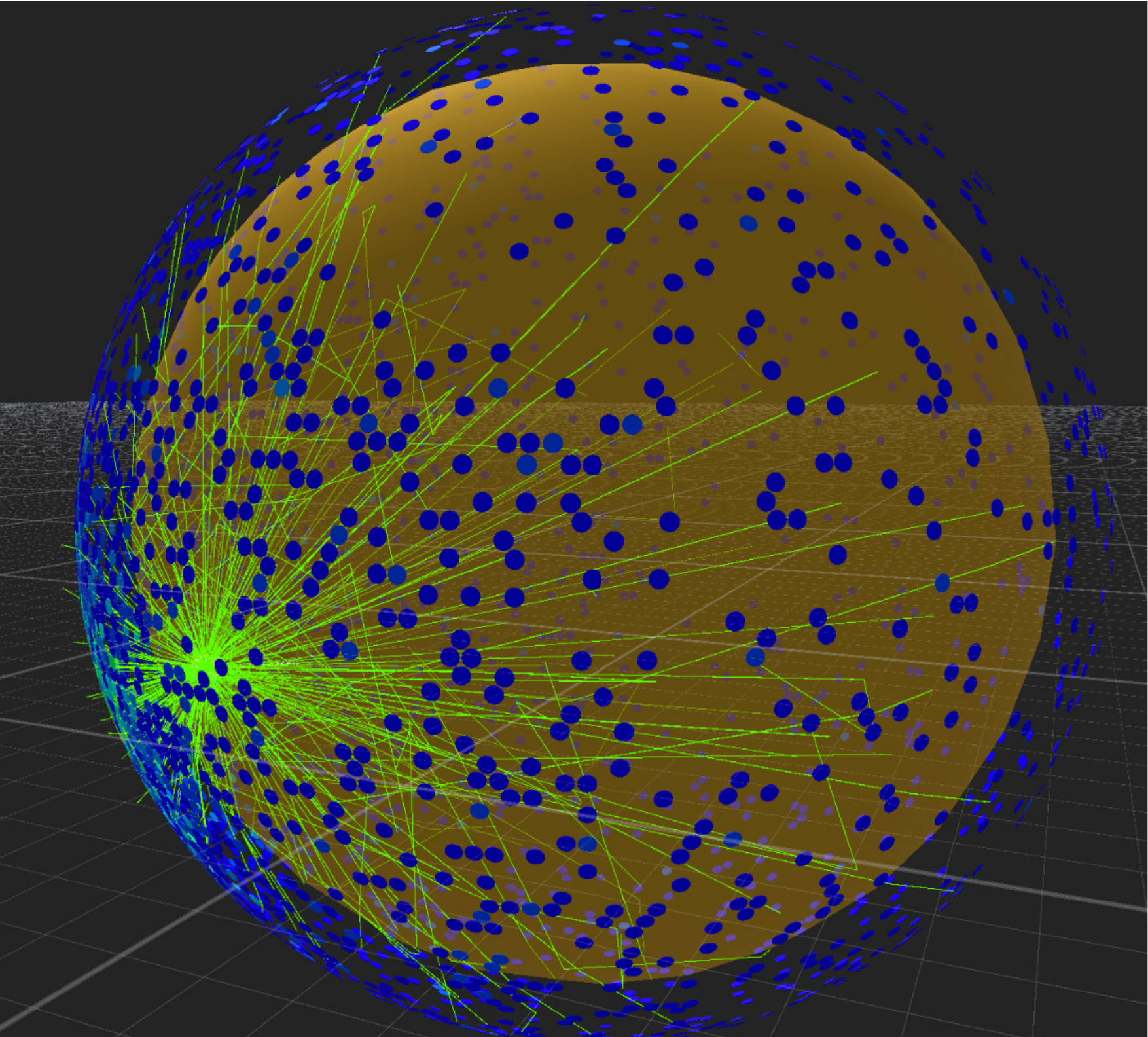}
\caption{\label{fig_photonTracks} Optical photon paths of a Monte Carlo event in JUNO Central Detector. 
The green lines show the true propagation paths of photons in simulation.}
\end{figure}  

As shown in Fig.~\ref{fig_photonTracks}, ELAINA also provides the function to draw the propagation paths of optical photons. 
When the energy of a particle is deposited in the LS, the yield photons do not always hit the PMTs in straight lines. 
Many optical processes may happen during propagation, such as Fresnel refraction, Fresnel reflection, absorption and re-emission, total internal reflection and Rayleigh scattering.
When the users are analyzing the simulation data with ELAINA, the optical paths with a specific type of optical process can be highlighted or selected in the visualization system so that the users can focus on the optical processes that they are interested in.

\subsection{Event information}
When observing the simulation and reconstruction events, it is not enough to only know the PMT hits distribution. 
The detail information of each PMT and the simple statistics of the whole event will be useful, which is also available in the visualization system.

As shown in Fig.~\ref{fig_hits}, on the main panel of ELAINA, the bottom right corner is a set of event control widgets.
The two histograms on the bottom left corner give the distributions of the first hit time and the total hits number on each PMT, 
from which the users will know how many PMT hits in a certain time interval of an event and how many PMTs receiving the same number of photons. 
Statistics of the current event is shown in the top left corner, including the total number of PMT hits in the event, the maximum hit time, etc. 
The top right corner gives the detailed information of the selected 3D object in the scene. 
For instance, users can check the geometry information of a specific PMT and its hit status in an event, such as its identifier, position, first hit time and the number of photons it received. 

\subsection{2D projection}
Besides the drawings in a 3D scene, 2D plots in event display are also useful for data analysis. 
In study of the reconstruction algorithm, some physicists may try to use the deep learning method to work on the event vertex and energy reconstruction problem. 
Many powerful and popular models in deep learning is based on the Convolutional Neural Network \cite{bib:CNN}, whose input data is usually in the format of 2D picture in some deep learning frameworks. 

\begin{figure}[htbp]
\centering 
\includegraphics[width=12cm]{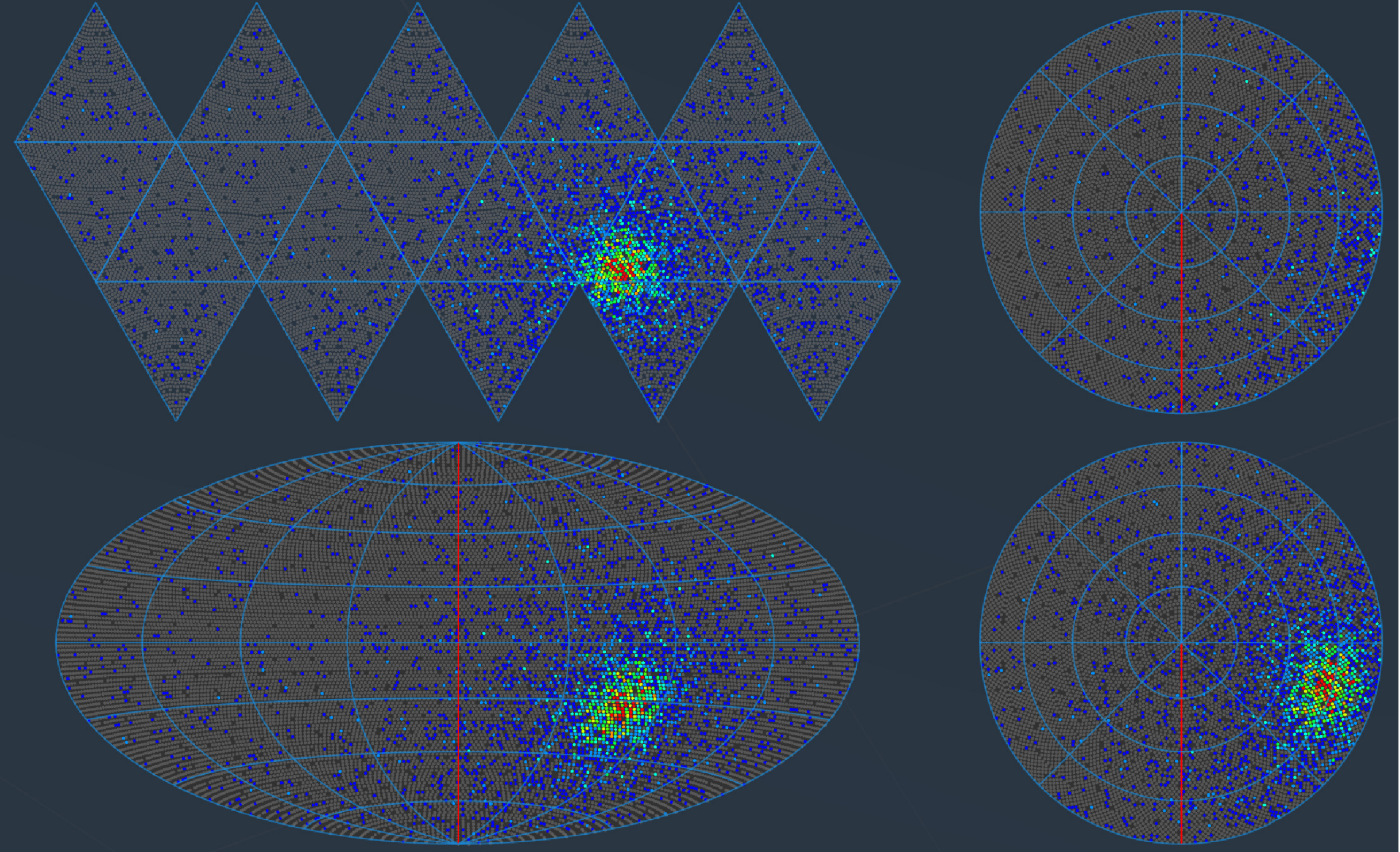}
\caption{\label{fig_2dProjection} 2D projection plots of the 20-inch PMTs in JUNO Central Detector. 
The top left is the Snyder Projection. 
The bottom left is the Hammer-Aitoff Projection. 
The right two plots are the Lambert Projection extended from the North Pole and South Pole respectively.}
\end{figure}

As shown in Fig.~\ref{fig_2dProjection}, ELAINA provides four kinds of 2D projection plots to draw the PMT hits distribution with three different equal-area projection methods for the PMTs arranged on the spherical surface of CD. 
The top left one uses the Snyder Projection \cite{bib:Snyder}, which transforms the surface of sphere into twenty triangles. 
The bottom left one is the Hammer-Aitoff Projection \cite{bib:Snyder}, which is widely used in many experiments. 
The other two on the right are the Lambert Projection \cite{bib:mapProjection} extended from the north pole and south pole of the sphere of CD.

\section{Features and advantages}

\subsection{Comparison with ROOT-based event display}
The Unity-based visualization system ELAINA has some distinct difference from the ROOT-based event display SERENA in JUNO. 
SERENA is developed with the ROOT EVE package and is integrated into the JUNO offline software. 
The strong integration with JUNO framework makes it convenient to directly load detector geometry data and the event data generated by offline software. 
The data analysis and the visualization both run in the same environment. 
The ROOT-based event display is often set up in the server, which usually provides powerful computing resources for data analysis but does not help much for the performance of visualization. 
To run the visualization remotely, users need to set up the client software.
The performance and user experience are also limited by the network connection.

ELAINA, the new detector and event visualization system based on Unity, is a specialized solution for users to run the event display locally. 
It is a client that is totally independent of the JUNO offline software. 
Different from the ROOT-based system, it requires to extract the detector and event data with JUNO offline macros into the text format that Unity can read. 
With the powerful multi-platform supports of Unity engine, the system can be built into applications for different platforms easily. 
So long as the extracted data is provided, the visualization can be implemented in the Unity engine with fancier visual effects.
The development with Unity is also flexible.
If the users' devices provide graphic acceleration power such as GPU, it can be fully utilized to improve the running and visualization performance of the Unity-based event display, which is also one of its advantages over the programs running remotely. 

\subsection{Multi-platform support}
ELAINA is a local running visualization system for JUNO.
Different users may use different operating systems on their personal computers.  
Linux is widely used in HEP community. 
However, some users like to use Windows and others may prefer to use MacOS. 
To allow the users to run the visualization locally, the applications working for different operating systems are needed. 
With the advantage of Unity, the Unity-based visualization system can be easily built into executable programs for different platforms so that the users can choose to download and install the matching program. 
We have built the ELAINA project into the applications for Windows, Linux and MacOS. 
It has been fully tested in Windows 8.1, Windows 10, Ubuntu 16.04, and MacOS 10.13. 
All applications have run successfully in the corresponding operating systems. 

\subsection{Potentials for future development}
The event display based on Unity is extensible. 
It has great potentials for further development. 
Besides the mainstream operating systems mentioned above, Unity can also build the project for mobile devices such as pad and cellphone.
It can even be built into the html format, which can run with the Internet browser such as Chrome and Firefox. 
A test shows that running the event display on the browser is feasible and the size of the website program can be reduced if the application is built with lower resolution and simpler visual effect. 
The Unity-based event display system is currently only developed for offline analysis, but it is not difficult to upgrade it for online event display and  onsite monitoring in the future.

To realize the visualization of event hits, basically what it needs to do is to match the event hits data with the corresponding detector elements. 
The implementation can be divided into two parts, building the detector and assigning the associated hits. 
The first and most important step is to visualize the detector structure, which means using the detector data to build the model in 3D space. 
Some HEP experiments use GDML to describe the structure of their detectors. 
It is possible to convert the GDML data into exported 3D file formats and then import the data into the Unity editor. 
As mentioned in previous section, there are two ways to initialize the detector structure. 
The simple way is always available if the developers have enough knowledge of the experiments. 
The other way with full geometry and structure transformation is available if the detector structure can be described in the generic 3D file format. 
It means that the Unity-based visualization system is not limited to JUNO only, but can be quickly implemented in other more complex experiments with some modifications.

With the advantage of Unity, the developers can easily build their projects for different platforms including the Virtual Reality platforms, which are developing rapidly recently. 
The VR version can provide much more vivid visualization for users and outreach of the experiment. 
Furthermore, the VR version of detector visualization has the potential to simulate the onsite installation situation and to  help optimizing the detector installation procedure.

\section{Conclusions}
A method of detector and event visualization based on Unity has been proposed.
It has been successfully implemented in the JUNO experiment to develop the event display software ELAINA. 
The method provides an intuitive way to visualize the detector, to observe and analyze events, and to tune the reconstruction algorithm. 
The Unity based event display software is independent of any offline software of an experiment, which makes it convenient for users to run the detector and event visualization on their local computers. 
It also has great potentials to develop more powerful functions and applications such as online monitoring and VR programs. 
With its successful application in JUNO, the Unity based event display system can also be implemented in other HEP experiments.

\acknowledgments

The authors would like to acknowledge the supports provided by the National Natural Science Foundation of China (11675275, 11405279, 11805294), the Strategic Priority Research Program of Chinese Academy of Sciences (XDA10010900) and the China Postdoctoral Science Foundation (2018M631013).



\begin{thebibliography}{99}

\bibitem{bib:Unity} W. Goldstone, \emph{Unity game development essentials}, Packt Publishing Ltd (2009).
\bibitem{bib:ROOT} R. Brun, F. Rademakers, \emph{ROOT - An object oriented data analysis framework}, \emph{Nucl. Instrum. Meth. A} {\bf 389} (1997) 81-86.
\bibitem{bib:EVE} M. Tadel, \emph{Overview of EVE -- the event visualization environment of ROOT}, \emph{J. Phys. Conf. Ser.} {\bf 219} (2010) 042055.
\bibitem{bib:ALICE} ALICE Collaboration et al., \emph{The ALICE experiment at the CERN LHC}, \emph{JINST} {\bf 3} (2008) S08002.

\bibitem{bib:CMS} CMS Collaboration et al., \emph{The CMS experiment at the CERN LHC}, \emph{JINST} {\bf 3} (2008) S08004.
\bibitem{bib:Fireworks}  D. Kovalskyi et al., \emph{Fireworks: A physics event display for CMS}, \emph{J. Phys. Conf. Ser.} {\bf 219} (2010) 032014.
\bibitem{bib:Atlantis} N. Konstantindis et al., \emph{The Atlantis event visualisation program for the ATLAS experiment}, \emph{Computing High Energy Physics Nucl. Phys.} {\bf 2004} (2004) 361.
\bibitem{bib:VP1} T. Kittelmann et al., \emph{The Virtual Point 1 event display for the ATLAS experiment}, \emph{J. Phys. Conf. Ser.} {\bf 219} (2010) 032012.
\bibitem{bib:ATLAS} G. Aad et al.,  \emph{The ATLAS experiment at the CERN large hadron collider}, \emph{JINST} {\bf 3} (2008) S08003.
\bibitem{bib:SketchUp} T. Sakuma, T. McCauley, \emph{Detector and Event Visualization with SketchUp at the CMS Experiment}, \emph{J. Phys. Conf. Ser.} {\bf 513} (2014) 022032.

\bibitem{bib:JUNO_CDR} T. Adam, F. An, G. An et al., \emph{JUNO conceptual design report}, arXiv:1508.07166.
\bibitem{bib:JUNO_physics} F. An, G. An, Q. An et al., \emph{Neutrino physics with JUNO}, \emph{J. Phys. G} {\bf 43}(3) (2016) 030401.

\bibitem{bib:SERENA} Z. You, K. Li, Y. Zhang et al., \emph{A ROOT based event display software for JUNO}, \emph{JINST} {\bf 13} (2018) T02002.

\bibitem{bib:XQuartz} \emph{XQuartz webpage}, \href{https://www.xquartz.org}{https://www.xquartz.org}.
\bibitem{bib:VNC} T. Richardson, Q. Stafford-Fraser et al., \emph{Virtual network computing} \emph{IEEE Internet Comput.} {\bf 2}(1) (1998) 33-38.
\bibitem{bib:TeamViewer} \emph{TeamViewer webpage}, \href{https://www.teamviewer.com}{https://www.teamviewer.com}.

\bibitem{bib:CAMELIA} \emph{CAMELIA webpage}, \href{http://medialab.web.cern.ch/content/camelia}{http://medialab.web.cern.ch/content/camelia}.
\bibitem{bib:PMT} A. G. Wright,  \emph{The Photomultiplier Handbook}, Oxford University Press (2017).
\bibitem{bib:Geant4} S. Agostinelli, J. Allison, K. Amako et al., \emph{GEANT4 - A simulation toolkit}, \emph{Nucl. Instrum. Meth. A} {\bf 506}(3) (2003) 250-303.
\bibitem{bib:GDML}R. Chytracek, J. McCormick, W. Pokorski and G. Santin, \emph{Geometry description markup language for physics simulation and analysis applications}, IEEE Trans. Nucl. Sci., 53: 2892 (2006)
\bibitem{bib:EDM} T. Li, X. Xia, X. Huang, J. Zou, W. Li, T. Lin et al., \emph{Design and development of JUNO event data model}, \emph{Chin. Phys. C} {\bf 41}(6) (2017) 066201.

\bibitem{bib:Blender} L. Flavell, \emph{Beginning Blender: Open Source 3D Modeling, Animation, and Game Design}, Apress, (2011).
\bibitem{bib:Maya} \emph{Maya webpage}, \href{https://www.autodesk.com/products/maya/overview}{https://www.autodesk.com/products/maya/overview}.
\bibitem{bib:3ds} \emph{3ds Max webpage}, \href{https://www.autodesk.com/products/3ds-max/overview}{https://www.autodesk.com/products/3ds-max/overview}.
\bibitem{bib:GDMLMethod}Z. Y. You, Y. T. Liang, Y. J. Mao, \emph{A method for detector description exchange among ROOT GEANT4 and GEANT3}, Chin. Phys. C, 32(7): 572-575 (2008)
\bibitem{bib:BesGDML} Y. Liang, B. Zhu, Z. You et al., \emph{A uniform geometry description for simulation, reconstruction and visualization in the BESIII experiment}, \emph{Nucl. Instrum. Meth. A} {\bf 603}(3), (2009) 325-327.
\bibitem{bib:CADMesh} C. M. Poole, I. Cornelius et al., \emph{A cad interface for geant4}, \emph{Australas. Phys. Eng. Sci. Med.} {\bf 35}(3) (2012) 329-334.
\bibitem{bib:FreeCAD} \emph{FreeCAD GDML module webpage}, \href{http://cad-gdml.in2p3.fr}{http://cad-gdml.in2p3.fr}.

\bibitem{bib:JUNO_Geometry}  K. Li, Z.You, Y. Zhang et al., \emph{GDML based geometry management system for offline software in JUNO}, \emph{Nucl. Instrum. Meth. A} {\bf 908} (2018) 43-48.

\bibitem{bib:CNN} A. Krizhevsky, I. Sutskever, G. E. Hinton, \emph{Imagenet classification with deep convolutional neural networks}, \emph{Adv. Neural Inf. Process. Syst.} (2012) 1097-1105.
\bibitem{bib:Snyder} J. P. Snyder, \emph{An equal-area map projection for polyhedral globes}, \emph{Cartographica: Int. J. Geogr. Inf. Geovis.} {\bf 29}(1) (1992) 10-21.
\bibitem{bib:mapProjection} Q. Yang, J. Snyder, and W. Tobler, \emph{Map projection transformation: principles and applications}, CRC Press (1999).











\end{thebibliography}
\end{document}